\documentclass{emulateapj}
\usepackage{amssymb,amsmath,graphicx,longtable,subfigure,wrapfig}
\usepackage{rotating}
\usepackage[colorlinks,linkcolor=blue,anchorcolor=green,citecolor=blue]{hyperref}
\usepackage{url}
\usepackage{natbib}
\usepackage{multibib}
\usepackage[english]{babel}
\usepackage{rotating}
\usepackage{xcolor}

\shorttitle{Remnant radio galaxy in A2065}
\shortauthors{Lal, D. V.}

\begin{document}


\title{The Discovery of a Remnant Radio Galaxy in A2065 Using GMRT}

\author{Dharam V. Lal}
\affil{National Centre for Radio Astrophysics - Tata Institute of Fundamental Research, Post Box 3, Ganeshkhind P.O., Pune 411007, India}
\email{dharam@ncra.tifr.res.in}

\begin{abstract}

The upgraded Giant Metrewave Radio Telescope (GMRT) has been used to map the cluster A2065 at $z$ = 0.0726.  We report the discovery of a remnant radio galaxy at the peripheral cluster region.  The spatially resolved radio emission from the remnant radio galaxy shows an elongated, bar-shaped structure, whose size is $\approx$ 52$^{\prime\prime}$ $\times$ 110$^{\prime\prime}$ ($\simeq$ 72 $\times$ 152 kpc$^2$).
Our study with the multiwavelength GMRT data and \textit{Chandra} data shows that across the remnant radio galaxy there is a hint of a surface-brightness edge in the hot X-ray gas.  We detect tentative flattening of the radio spectral index as the old plasma at the near end of the surface-brightness edge is reinvigorated by the passage of possible shock front and shows the expected change in radio emission characteristics.
We suggest that the remnant radio galaxy has been seeded by the lobes of the active galactic nucleus (AGN), hosted by the WISEA J152228.01$+$274141.3 source, demonstrating the connection between AGNs and remnant radio sources.
Although the number of known remnant radio sources is beginning to increase, we emphasize the need for better data to understand the physics and nature of poorly understood remnant radio sources.
\end{abstract}


\section{Introduction}
\label{intro}

The hierarchical structure formation scenario suggests that cluster mergers are the mechanisms by which galaxy clusters form and these mergers are the most energetic phenomena in the universe since the Big Bang.
A large number of clusters of galaxies are known to contain large-scale, low-surface-brightness diffuse radio emission whose origin is not often related to the active galactic nuclei (AGNs) but to the intracluster medium (ICM).  According to the location with respect to the cluster center, these sources are of two key types, radio halos and relics.  Radio halos show a regular, amorphous emission around the cluster center of a size from a few hundred kiloparsecs (minihalo) to megaparsecs (halo), whereas radio relics show an irregular emission located in peripheral cluster regions of a size from several hundreds of kiloparsecs to megaparsecs.
Another third broad type of sources comprises of revived fossil plasma sources and phoenices.  These do not have any preferred location in the ICM are of size from several tens of kpc to a few hundred kpc and trace AGN radio plasma that has somehow been reenergised through the processes in the ICM \citep{vanWeerenetal2019,Kempneretal,GiovanniniandFeretti2002}.

The radio jets emanating from an AGN arise from the accretion onto a supermassive black hole.
These form synchrotron-emitting radio lobes in a radio-loud AGN in the environments of their host galaxies.
The active phase of an AGN can last several tens of megayears, following this the radio jets stop and their features, namely an unresolved radio core, radio jets, hotspots, and radio lobes, all start to fade away due to synchrotron cooling.  This phase once the jets have switched off is called as the remnant phase of a radio galaxy, or merely the remnant radio galaxy.
The remnant radio galaxy falls into the third broad type of sources in the ICM, which is poorly understood and presently very few sources are known \citep{Quicietal,Brienzaetal,Parmaetal2007}.

The high spatial resolution of the \textit{Chandra} and upgraded Giant Metrewave Radio Telescope (GMRT) allow us to study the interplay between X-ray-emitting hot gas and nonthermal radio emission.
We have obtained new upgraded GMRT data, and have used archive GMRT and \textit{Chandra} data of the nearby $z$ = 0.0726 \citep{1999ApJS..125...35S} richness class 2 cluster, A2065, which is one of the 10 galaxy clusters that make up the Corona Borealis supercluster.
\citet{Postmanetal} noted that two cD galaxies dominate the cluster center in the optical and their radial velocities differ by $\sim$600 km~s$^{-1}$.

\begin{table*}
\caption{Summary of the radio observations and image properties.}
\begin{center}
\begin{tabular}{clcccccccc}
\hline\hline
Obs. Band & Obs\_ID  &  Obs. Date & Flux Cal. & Phase Cal. & $\nu$ & $\Delta\nu$ & $t_{\rm int.}$ & FWHM, P.A. & \textsc{rms}  \\
   & &      &  & &  (MHz) &   (MHz) & (hr) & ($^{\prime\prime}\times^{\prime\prime}, ^{\circ}$)& (mJy~beam$^{-1}$)              \\
 (1) &  \multicolumn{1}{c}{(2)} & (3) & (4) & (5) & (6) & (7) & (8) & (9) & (10) \\
\hline\noalign{\smallskip}
 \multicolumn{10}{l}{``classic" GMRT} \\
150 MHz & 12RKA02  & 2007 Aug 10 & 3C\,286 & 1459$+$716 & 152  &  ~6  &  0.9 &26.90$\times$15.19, 136.15 & 4.98 \\
240 MHz & 14RMD01  & 2008 Jul 26 & 3C\,286 & 1513$+$236 &  240  &  ~8  &  5.4 &13.95$\times$10.94, ~98.89 & 0.56 \\
610 MHz & 14RMD01  & 2008 Jul 26 & 3C\,286 & 1513$+$236 &  606  &  16  &  5.4 &~4.93$\times$3.97, ~65.80 & 0.06 \\
 \multicolumn{10}{l}{Upgraded GMRT} \\
250--500 MHz & 32\_084  & 2017 Jun 27 & 3C\,147 & 3C\,286 & ~400  & 200  &  1.3 &~9.48$\times$6.38, ~69.18 & 0.06 \\
1050--1450 MHz & 32\_084& 2017 Jun 28 & 3C\,147 & 1609$+$266 & 1274  & 400  &  0.9 &11.13$\times$5.90, ~95.45 & 0.03 \\
\hline
\end{tabular}
\end{center}
\label{tab:obs-log-prop}
\tablecomments{Each column, in ascending order, details the observing band (Column 1), project code (Column 2), the date observations were conducted (Column 3), the primary flux density calibration source observed (Column 4), the secondary phase calibration source observed (Column 5), the central frequency of the receiver band (Column 6), the bandwidth available within each band (Column 7), the approximate time spent in each observing run (Column 8), the shape of the restoring beam (Column 9), and the \textsc{rms} noise at the half-power point (Column 10).}
\end{table*}

Previous radio study using 100~m Green Bank Telescope at 1.4\,GHz \citep{Farnsworthetal} detected a smooth diffuse structure $\simeq$1~Mpc in extent, a giant radio halo, whereas a recent study showed a slightly smaller extent for the radio halo using Low-Frequency Array (LOFAR) at 153\,MHz \citep[in][]{Drabentetal}.
Similarly, previous X-ray studies led to detections of two surface-brightness peaks coincident with the two cD galaxies observed in the cluster's center.
The data indicated that the cluster is an unequal-mass merger and only one of the two cooling cores has survived the merger.  The northern cluster seems to have fallen into the more massive southern cluster from the southeast, and having lost its gas content, it has now moved to its present location to the northwest.  Evidence of shock with $M$ $\approx$ 1.7 at $\sim$140~kpc from the southern cD galaxy too is noted and the cooling flow is displaced to the southeast of the southern cD galaxy \citep{Chatzikosetal2006,Markevitchetal1999}.  In this study, we report the discovery of a remnant radio galaxy in cluster A2065 using the GMRT.  We also present an X-ray analysis of archive \textit{Chandra} data, which enhances the interpretation of the current discovery.

We define spectral index, $\alpha$ as, $S_\nu$ $\propto$ $\nu^{\alpha}$, where S$_\nu$ is the flux density at frequency, $\nu$.
To estimate the intrinsic parameters, we use a $\Lambda$CDM cosmology with $\Omega_{\rm m}$ = 0.27, $\Omega_{\Lambda}$ = 0.73, and H$_0$ = 70 km s$^{-1}$ Mpc$^{-1}$.  At the redshift of the cluster, 1~arcsec corresponds to 1.384~kpc.
All uncertainties are at 1$\sigma$ error bars unless otherwise stated.
The position angles (P.A.s) are measured from north to east.  Throughout, positions are given in J2000 coordinates.

The organization of the paper is as follows.
We present the X-ray and radio observations and their data analyses in Sec.~\ref{radio-x-ray-data}.
Our results, {\it i.e.}, radio morphology, spectral structure, and X-ray gas properties are presented in Sec.~\ref{sec.results}.
In Sec.~\ref{sec.discuss}, we discuss our observational results, source diagnostics, and dynamical interpretation of the remnant radio galaxy.
Sec.~\ref{sec.sum-conc} summarizes our conclusions.

\section{Observations}
\label{radio-x-ray-data}

\subsection{Chandra data}

A2065 was observed by \textit{Chandra} with the \textsc{acis-i} detector on 2002 August 18 and November 24 for 28 and 22~ks (Obs\_ID 3182) respectively, and on 2007 September 13 for 5~ks (Obs\_ID 7689).  The data reduction of these three archival data sets has been performed independently.  We used the calibration databases, \textsc{caldb} version 4.7.6 and \textsc{ciao} version 4.9.
After the initial calibration, {\it i.e.}, standard filtering and removal of bad pixels, the event files of the three observations were projected to a common reference point and then merged to create Gaussian-smoothed, background-subtracted, exposure-corrected image in the broadband 0.5--7.0~keV energy range.

We excised the central AGN and background point sources.  Using the \textsc{ciao} tool \textsc{specextract}, we extracted the source, background spectra, and the response files of selected regions in the \textit{Chandra} images separately for the three data sets.
The spectra for each data set were grouped so that each bin contained at least 20 counts.  We restricted our spectral extraction in the energy range 0.5--7.5~keV, and fitted for the thermal gas emission with \textsc{xspec} \citep[version 12.10.1,][]{Arnaud1996} using the isothermal \textsc{apec} model.
The abundance was fixed to 0.3 $\times$ solar \citep{Chatzikosetal2006} and all spectral fits include absorption by the Galactic column \citep[= 2.9 $\times$ 10$^{20}$ cm$^{-2}$,][]{DickeyLockman}.

\begin{figure}[ht]
\begin{center}
\begin{tabular}{c}
\includegraphics[width=8.45cm]{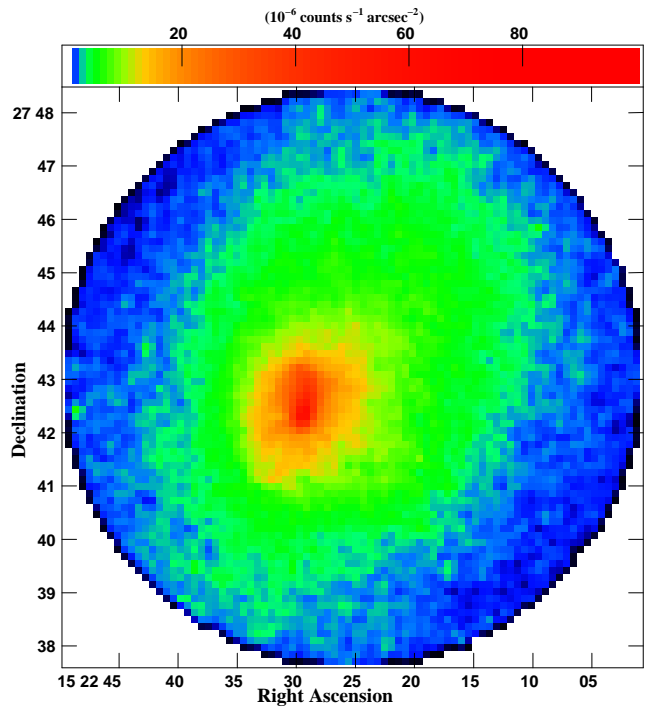}
\end{tabular}
\caption{Exposure-corrected, background-subtracted, broadband (0.5--7.0~keV) \textit{Chandra} image of A2065.  The image was binned (bin = 16) and smoothed using a Gaussian kernel of 10$^{\prime\prime}$.  The color bar shows the surface-brightness, in 10$^{-6}$ counts~s$^{-1}$ per 8 $\times$ 8 pixel binning.  The X-ray peak is located at R.A. 15:22:29.18, decl. 27:42:21.05.  The peak surface-brightness and the faintest regions transitioning between blue and green have a surface-brightness of 6.0 $\times$ 10$^{-5}$ counts~s$^{-1}$~arcsec$^{-2}$ and 1.6 $\times$ 10$^{-6}$ counts~s$^{-1}$~arcsec$^{-2}$ respectively.
The surface-brightness is displayed in logarithmic scales to emphasize low-surface-brightness emission.}
\label{fig:f1}
\end{center}
\end{figure}

\subsection{GMRT data}
\label{uGMRT-data}

\begin{figure*}[ht]
\begin{center}
\begin{tabular}{cc}
\includegraphics[width=17.8cm]{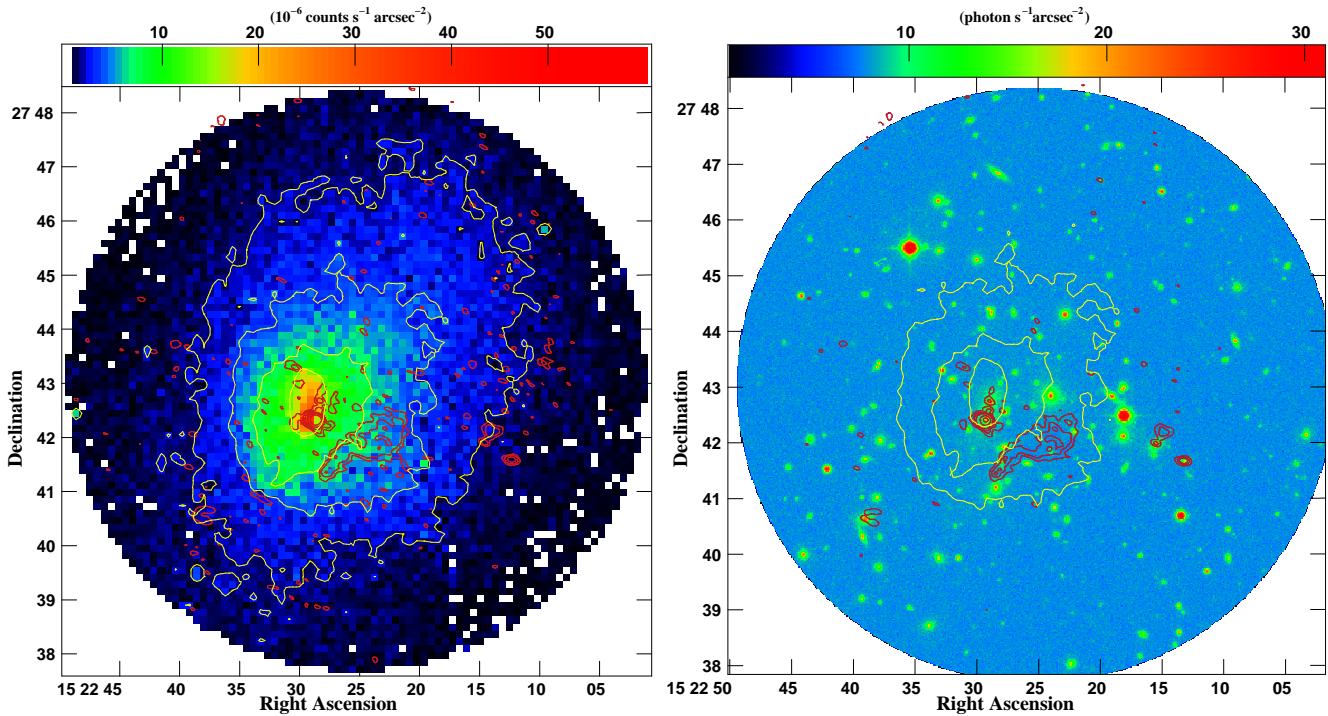}
\end{tabular}
\caption{\textit{Chandra} X-ray contours (see Figure~\ref{fig:f1}) and 250--500 MHz band radio contours overlaid on the
exposure-corrected, background-subtracted broadband (0.5--7.0~keV) \textit{Chandra} image (left panel) and Digitized Sky Survey image (right panel) of A2065.
The yellow X-ray contour levels are 3.18, 6.36, 12.72 and 25.44 $\times$ 10$^{-6}$ counts~s$^{-1}$~arcsec$^{-2}$
and the red radio contour levels are $-$3, 3, 6, 12, 24 and 48 $\times$ 77.5 $\mu$Jy~beam$^{-1}$ on the \textit{Chandra} image (left panel).
The yellow X-ray contour levels are 3.18, 6.36 and 12.72 $\times$ 10$^{-6}$ counts~s$^{-1}$~arcsec$^{-2}$
and the red radio contour levels are $-$3, 3, 6, 9, 15, 42 and 84 $\times$ 0.1 mJy~beam$^{-1}$ on the Digitized Sky Survey image (right panel).
The radio emission from the two cD galaxies (the brighter southern galaxy 2MASS J152229.17$+$274227.5 and the fainter northern galaxy 2MASX J152228.92$+$274244.1) of the cluster is shown on the east of the remnant radio galaxy emission (denoted with two blue plus marks).  The emission from the remnant radio galaxy corresponds to the 19.19 mag, WISEA J152228.01$+$274141.3 source, denoted by a red plus mark that is 7\farcs8 away from V1CG\,136 galaxy located at R.A.: 15:22:28.062, decl.: $+$27:41:41.51, $z$ = 0.072 \citep{2014ApJ...795L..21G} denoted by a black cross.
The color scale was chosen to emphasize the X-ray surface-brightness edge near the remnant radio galaxy.
The images are displayed in logarithmic scales to emphasize a faint optical host associated with the remnant radio galaxy.  Note the X-ray peak surface-brightness seems to be associated with the radio peak corresponding to the southern cD galaxy.}
\label{fig:f2}
\end{center}
\end{figure*}

A summary of the radio observations is presented in Table~\ref{tab:obs-log-prop}.
We inspected the data for bad antennas, bad time stamps and data impacted by radio frequency interference. 
The data were analyzed using the standard reduction methodologies \citep[see also][]{Lal2020}.
After the initial steps of flux density and bandpass calibration, all corrupted data were removed using standard routines.
The bandpass calibrated data for wide-bandwith GMRT data, {\it i.e.}, the 250--500 MHz band and 1050--1450 MHz band data were split into five 40~MHz (from 300--500 MHz) and eight 50~MHz (from 1050--1450 MHz) subbands respectively.  Subsequently, each subband data was analyzed separately, similar to narrowband ``classic" GMRT data.
Several cycles of phase-only self-calibration and imaging were performed and the calibration solutions were applied to the target data to correct for residual phase errors.  To take care of the wide-field imaging at low GMRT frequencies, we performed faceting imaging.
Problematic bright sources in the field of view whose side lobes affected the area of the target were subtracted, a.k.a. peeled; this (direction-dependent) calibration step improved the dynamic range of images. The calibrated data for five 40~MHz 250--500 MHz band data and eight 50~MHz 1050--1450 MHz band data were joined to form full 200~MHz and 400~MHz, respectively, wide-bandwidth-calibrated visibility data sets.  The combined wide-bandwidth-calibrated data (and similarly the narrowband-calibrated ``classic" GMRT data) were imaged using the \textsc{casa} task \textsc{tclean}.
We used 3D imaging (gridder = `widefield'), two Taylor coefficients (nterms = 2, tt0 and tt1) and Briggs weighting (robust = 0.5) in task \textsc{tclean}.  A final amplitude-and-phase self-calibration with solution interval equal to the length of observation was then carried out using the \textsc{gaincal} and \textsc{applycal} tasks in \textsc{casa}.
The two Stokes polarizations, RR and LL, were combined to obtain the final total intensity Stokes I image, and corrected for the GMRT primary beam response.
This provides images at center frequencies (tt0) and the corresponding in-band spectral index maps (tt1/tt0).
The amplitude errors are estimated to be within 5\% or less for archival narrow-bandwidth and wide-bandwidth GMRT data.
Table~\ref{tab:obs-log-prop} also provides image properties, including the restoring beams and root mean square (\textsc{rms}) noise levels at half-power points of the final images.
The uncertainty in the flux density measurements are estimated as
$$
\left[(S_\nu \times f)^2 + (\textsc{rms})^2 \times N_{\rm beams}\right]^{0.5},
$$
where $S_\nu$ is the flux density, $f$ is an absolute flux density calibration uncertainty, \textsc{rms} is the noise, and $N_{\rm beams}$ is the number of synthesized beams.
We use these error estimates for the flux density measurements when computing the error estimates for spectral index measurements.

\begin{figure}[ht]
\begin{center}
\begin{tabular}{c}
\includegraphics[width=8.45cm]{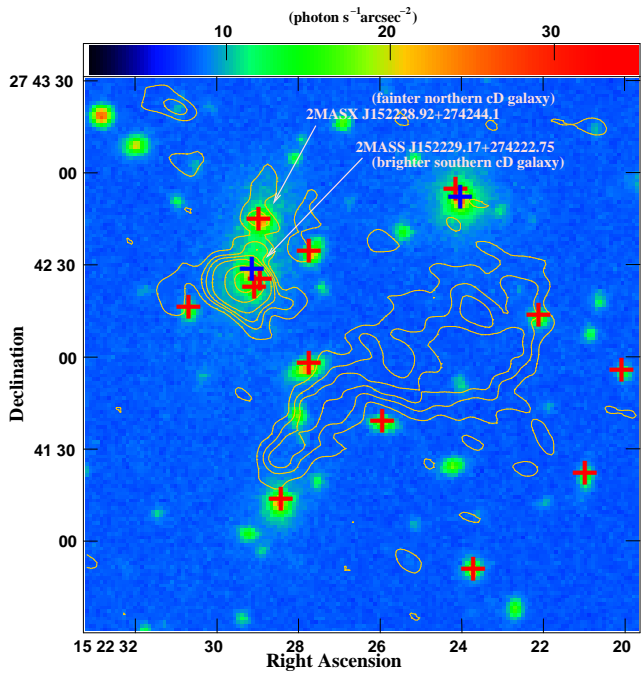}
\end{tabular}
\caption{The central 3$^{\prime}$\,$\times$\,3$^{\prime}$ cluster region with 250--500 MHz band radio contours overlaid on the Digitized Sky Survey image of A2065.   Note the northern cD galaxy registers well with the northern radio peak, but the southern cD galaxy is displaced by $\sim$2\farcs4 north of the radio peak.  The X-ray peak at the cluster center is coincident with the southern radio peak (see also Figure~\ref{fig:f2}).
The radio emission from the two cD galaxies (the brighter southern galaxy 2MASS J152229.17$+$274227.5 and the fainter northern galaxy 2MASX J152228.92$+$274244.1) of the cluster is shown on the east of the remnant radio galaxy emission.
The radio contour levels are $-$3, 3, 6, 9, 15, 42 and 84 $\times$ 80 $\mu$Jy~beam$^{-1}$.
The cluster-member galaxies are marked \citep[NED positions, and ``diffuse radio emission in the Corona Borealis supercluster field",][]{Drabentetal} and a majority ($\approx$60\%) show radio emission.  The two bright cluster galaxies, SDSS $+$230.6$+$27.7$+$0.08 (R.A.: 15:22:24.0, decl. $+$27:42:51; $z$ = 0.0723) and 2MASS J152229.17$+$274222.75 ($z$ = 0.074875) are marked in blue colour \citep{Postmanetal}.
}
\label{fig:f3}
\end{center}
\end{figure}

\section{Results}
\label{sec.results}

The exposure-corrected background-subtracted X-ray binned (bin = 16) and Gaussian-smoothed (10$^{\prime\prime}$ kernel) image in the 0.5--7.0~keV band is shown in Figure~\ref{fig:f1}.
Figure~\ref{fig:f2} shows \textit{Chandra} X-ray contours from Figure~\ref{fig:f1} and 250--500 MHz band radio contours overlaid on the exposure-corrected, background-subtracted broadband (0.5--7.0~keV) \textit{Chandra} (bin = 8) image (left panel) and Digitized Sky Survey image (right panel) of A2065.
The central 3$^{\prime}$\,$\times$\,3$^{\prime}$ cluster region with 250--500 MHz band radio contours overlaid on the Digitized Sky Survey image of A2065 is shown in Figure~\ref{fig:f3}.
Figure~\ref{fig:f3} also marks the cluster member galaxies, including the two bright cluster galaxies in marked in blue colour, SDSS $+$230.6$+$27.7$+$0.08 (NED; R.A.: 15:22:24.0, decl. $+$27:42:51; $z$ = 0.0723, $m$ = 16.1) and 2MASS J152229.17$+$274222.75 \citep[NED; $z$ = 0.074875, $m$ = 14.65][]{Postmanetal}, and a majority of them, $\approx$60\%, show radio emission.

Figure~\ref{fig:f4} shows clear detections for the southern cD galaxy and remnant radio galaxy at the GMRT bands, namely the 240 MHz band (top-left panel), 250--500 MHz band (top-right panel), 610 MHz band (bottom-left panel), and 1050--1450 MHz band (bottom-right panel).
The northern cD galaxy is only detected in our 250--500 MHz band data.
The images have \textsc{rms} noise values in the range 0.56--0.06~mJy~beam$^{-1}$ at the half-power points.  The 150 MHz band and 1050--1450 MHz band data sets had very short on-source integration times, $\lesssim$0.9~hr, and therefore these images have relatively high \textsc{rms} noise values (see Table~\ref{tab:obs-log-prop}).
Note that we also made images at several angular resolutions, including imaging data sets using the same range of baseline lengths, and do not find the presence or absence of additional morphological features.  The integrated flux densities computed thus are also consistent, within error bars with those reported in Table~\ref{tab:obs-log-prop}.

\subsection{Southern and northern cD galaxies}
\label{sec.Xray-two-cDs}

\begin{figure*}[ht]
\begin{center}
\begin{tabular}{c}
\includegraphics[width=17.8cm]{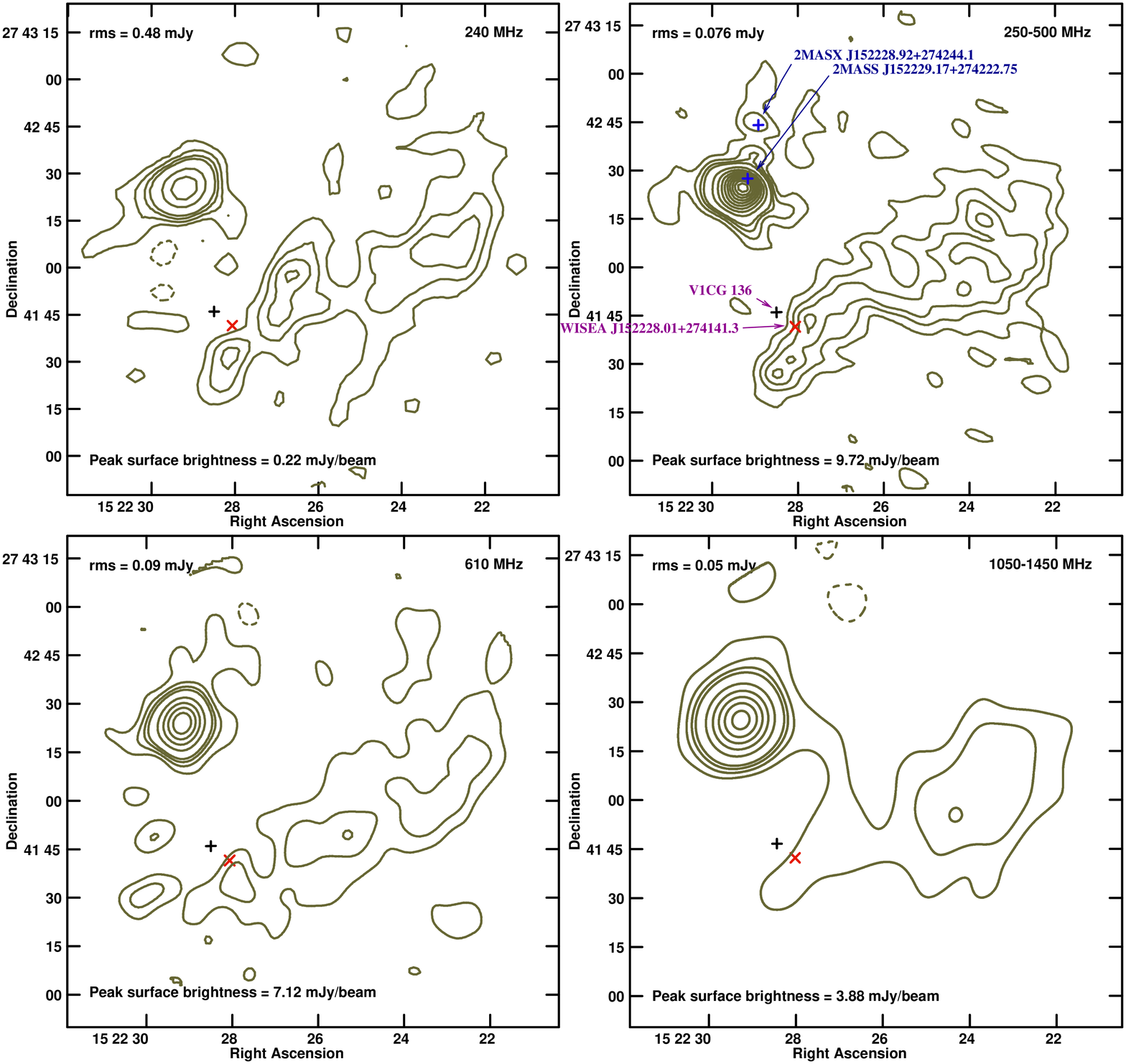}
\end{tabular}
\caption{Images of the cD galaxy and the remnant radio galaxy are shown using the GMRT bands, {\it i.e.}, 240 MHz band (top-left panel), 250--500 MHz band (top-right panel), 610 MHz band (bottom-left panel), and 1050--1450 MHz (bottom-right panel) data.
Two peaks (marked as blue cross-marks in the 250--500 MHz band image), the brighter southern and the fainter northern peaks correspond to the two cD galaxies, 2MASS J152229.17$+$274222.75 and 2MASX J152228.92$+$274244.1, respectively.
The radio contour levels plotted in each panel are $-$3, 3, 6, 9, 12, 24, 36, 48, 60 and 72 times the \textsc{rms} noise (denoted at upper-left corner in each panel) in units of mJy~beam$^{-1}$.
The black plus and the red cross marks are, respectively, the positions of the WISEA J152228.01$+$274141.3 source and optical host galaxy V1CG\,136 \citep[R.A.: 15:22:28.062, decl.: $+$27:41:41.51, $z$ = 0.072;][]{2014ApJ...795L..21G}.}
\label{fig:f4}
\end{center}
\end{figure*}

Figure~\ref{fig:f1} shows a diffuse X-ray tail extending to the north at the center and that the cluster emission is elongated along northwest-southeast direction.
The hot gas emission (i) shows a compact, bright region and a diffuse tail extending to the north at the center, and (ii) is extended along the northwest--southeast direction showing pieces of evidence for surface-brightness edges. 
Though the X-ray peak emission and peak of the radio emission register well (see Figure~\ref{fig:f2}, left), the position of the optical host of the southern cD galaxy (R.A.: 15:22:29.17, decl.: $+$27:42:27.5) is displaced by $\sim$2\farcs4 from the compact, bright, X-ray peak emission at the cluster center (see Figure~\ref{fig:f2}, right, and also Figure~\ref{fig:f3}).  A low-surface-brightness diffuse tail, both in X-ray as well as in radio images, extending to the north at the center of the image, reaches out to the second cD galaxy (R.A.: 15:22:28.92, decl.: $+$27:42:44.1; Figures~\ref{fig:f3} and \ref{fig:f4}, top-right panels).  The X-ray peak corresponding to the second northern cD galaxy and the peak of the second weaker component of radio emission at this location also register well.  The southern cD galaxy seems to have retained a cool core residing at the center of its subcluster and there is no evidence for a cool core corresponding to the northern cD galaxy \citep{Chatzikosetal2006,Markevitchetal1999}.  The closely associated radio and X-ray morphology at the center and its connection with the southern cD galaxy suggest that both X-ray and radio emissions are due to ram pressure stripping as the southern cD galaxy moves relative to the ICM.

The two blue plus signs in the 250--500 MHz band image marked in Figure~\ref{fig:f4} (top-right panel) correspond to the hosts of the brighter southern cD galaxy, 2MASS J152229.17$+$274222.75 at $z$ = 0.074875 and fainter northern cD galaxy, 2MASX J152228.92$+$274244.1 at $z$ = 0.072854 (see also Figure~\ref{fig:f3}).
The northern cD galaxy is only detected in our 250--500 MHz band image.
However, we not only detected the southern cD galaxy in our long observations, we also detected the peak radio emissions at the location of the southern cD galaxy at R.A.: 15:22:29.17, decl.: $+$27:42:27.5 in our short, 150 MHz band, and 1050--1450 MHz band data sets.
The two cD galaxies are barely resolved using LOFAR at 153~MHz \citep[``diffuse radio emission in the Corona Borealis supercluster field",][]{Drabentetal}.
The southern cD galaxy also had a 2.8$\sigma$ (upper limit) detection at 74\,MHz \citep[VLA Low-Frequency Sky Survey, VLSS:][]{2007AJ....134.1245C}.
It seems that around the southern cD radio galaxy there is some extended radio emission, which is more clearly seen in our 250--500 MHz band image (Figure~\ref{fig:f4}, top-right panel).  Since several cluster-member galaxies also show radio emission, it is unclear if this extended radio emission is indeed associated with the southern cD radio galaxy or with the two adjacent cluster-member galaxies southeast of it (see Figure~\ref{fig:f3}).

\begin{table*}[ht]
\tabletypesize{\scriptsize}
\caption{Radio flux densities for the remnant radio galaxy and the southern cD galaxy}
\begin{center}
\begin{tabular}{lccccccc}
\hline\hline
 & $S_{\rm 74\,MHz}$ & $S_{\rm 150\,MHz}$ & $S_{\rm 240\,MHz}$ & $S_{\rm 400\,MHz}$ & $S_{\rm 610\,MHz}$ & $S_{\rm 1250\,MHz}$ & $S_{\rm 1400\,MHz}$ \\
   \multicolumn{7}{c}{(mJy)} \\
\multicolumn{1}{c}{(1)} &  (2) & (3) & (4) & (5) & (6) & (7) & (8) \\
\hline\noalign{\smallskip}
Remnant radio galaxy & & & 46.52 $\pm$0.90 & 26.76 $\pm$0.08 & 2.60 $\pm$0.22  & 0.73 $\pm$0.15 &  \\
Southern cD galaxy & 395.2 $\pm$141.1 & 46.52 $\pm$0.90 & 36.26 $\pm$0.90 & 21.97 $\pm$0.08 & 11.12 $\pm$0.22 & 3.83 $\pm$0.09 & 3.09 $\pm$0.15 \\
\hline
\end{tabular}
\end{center}
\label{tab.rad-spectra}
\tablecomments{The total flux densities for the remnant radio galaxy were determined from integration in polygons with measured in areas encompassing the remnant radio galaxy,
whereas the peak flux density for the remnant radio galaxy is reported from our low-resolution, 15$^{\prime\prime}$ image at the 1050--1450\,MHz band (Column 7) shown in Fig.~\ref{fig:f4} (bottom-right panel)  (see also Sec.~\ref{sec.rad-spec} for a discussion). \\
The 2.8$\sigma$ upper limit at the 74\,MHz (Column~2),
peak flux density at 150 MHz band (Column~3),
measurements at 240\,MHz (Column~4), 400\,MHz (Column~5), 610\,MHz (Column~6) and 1250\,MHz (column~7), and
the data at 1400\,MHz (column~8) for the southern cD galaxy, are from the VLSS at 80$^{\prime\prime}$ angular resolution
\citep{2007AJ....134.1245C},
GMRT 150 MHz band, GMRT 240\,MHz band, upgraded GMRT 250--500\,MHz band, GMRT 610\,MHz band, upgraded GMRT 1050--1450\,MHz band data, and
\citet{Chatzikosetal2006}, respectively.}
\end{table*}

\subsection{Remnant radio galaxy}
\label{sec.radio-relic}

Figures~\ref{fig:f3} and~\ref{fig:f4} also suggest that the remnant radio galaxy is hosted by the 19.19 magnitude WISEA J152228.01$+$274141.3 source that is 7\farcs8 away from the optical galaxy V1CG\,136 (R.A.: 15:22:28.5, decl.: $+$27:41:46) at redshift $z$ = 0.072 \citep{2014ApJ...795L..21G}.  The cluster contains gas not only with an extremely wide range of temperatures and complex structure but that also shows a cold front at 30$^{\prime\prime}$ ($\simeq$ 41.5~kpc) and a shock-like bow-shaped feature beyond $\sim$100$^{\prime\prime}$ ($\simeq$ 140~kpc) on the southeast of the southern cD \citep{Chatzikosetal2006}.  Figure~\ref{fig:f2} shows that the northeastern boundary of the remnant radio galaxy coincides with the 1.3 $\times$ 10$^{-7}$ counts~s$^{-1}$~arcsec$^{-2}$ surface-brightness contour level; gas in false color transitioning from the green region to the blue region at a distance of $\sim$33$^{\prime\prime}$ ($\simeq$ 45.7~kpc) from the southern cD galaxy.
All four panels of Figure~\ref{fig:f4} suggest that the radio emission from the remnant radio galaxy has an elongated, bar-shaped structure.
The remnant radio galaxy, buried in the radio halo emission that is barely detected using LOFAR at 153~MHz seems to possess an elongated, bar-shaped structure \citep[``diffuse radio emission in the Corona Borealis supercluster field,][]{Drabentetal}, consistent with our results.
This suggests that the remnant radio galaxy has a steep spectrum, and its consistency in radio morphology gives us additional confidence in our image fidelity, imaging, calibration and data reduction processes.
The best sensitivity image for the low surface-brightness was obtained from the 250--500 MHz band data and we now discuss below the morphology of the remnant radio galaxy.

The remnant radio galaxy broadly consists of a high-surface-brightness southeastern part and a fainter low-surface-brightness extended northwestern part.  The two oppositely directed radio jets emanating from the apex of the WISEA J152228.01$+$274141.3 source initially traverse toward the southeast and northwest directions.  As the galaxy plows through the dense intracluster gas, these jets traversing in opposite directions form a trail, after sharp bends in the jets, behind the WISEA J152228.01$+$274141.3 source due to interaction with the ICM.  The width of the remnant radio galaxy increases from $\sim$17$^{\prime\prime}$ ($\simeq$ 23.5~kpc) at the southeastern end to $\sim$27$^{\prime\prime}$ ($\simeq$ 37.4~kpc) and reduces again before flaring to reach a maximum width of $\sim$35$^{\prime\prime}$ ($\simeq$ 48.4~kpc) and finally fading completely.  The average width of the narrower southeastern part is a factor of $\sim$2.3 smaller than the broader northwestern part.  It seems that a pinch is present at $\sim$35$^{\prime\prime}$ ($\simeq$ 48.4~kpc) of the $\sim$16\farcs2 ($\simeq$ 22.4~kpc) width from the southeastern end,  {\it i.e.}, between the two narrower and the broader parts.  The northeastern boundary toward the two cD galaxies of the remnant radio galaxy is sharp, while the emission fades more slowly at the southwestern boundary.
The overall source size, assuming elongation, of the bar-shaped structure is $\approx$ 52$^{\prime\prime}$ $\times$ 110$^{\prime\prime}$, corresponding to 72 $\times$ 152 kpc$^2$ at the assumed distance.

\begin{figure}[b]
\begin{center}
\begin{tabular}{c}
\includegraphics[width=8.45cm]{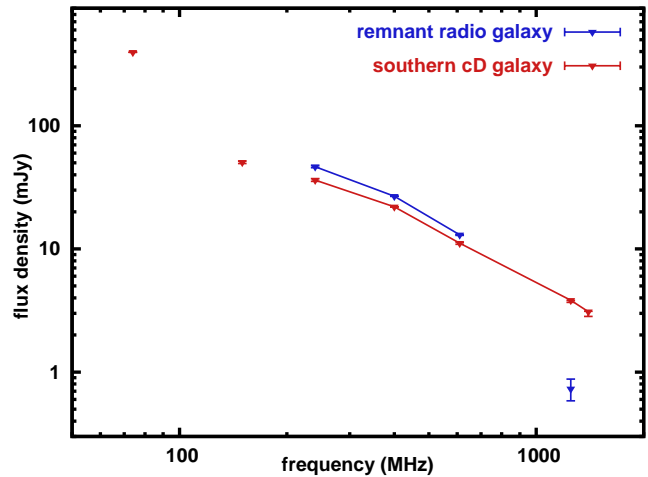}
\end{tabular}
\caption{Integrated radio spectrum of the remnant radio galaxy using the data in Table~\ref{tab.rad-spectra}.
Peak flux density is reported from our low-resolution, 15$^{\prime\prime}$ image at the 1050--1450\,MHz band for the remnant radio galaxy (see also Sec.~\ref{sec.rad-spec} for a discussion).
We also show the spectra of the southern cD galaxy for comparison; note that the (nearly) straight spectra of it provides a good check of our calibration.  A 2.8$\sigma$ limit is reported at the 74\,MHz \citep[VLSS;][]{2007AJ....134.1245C}, peak flux densities at the 150 MHz band, and 1050--1450\.MHz band are our measurements; the 1400\,MHz data is from the VLA FIRST survey \citep{Chatzikosetal2006} for the southern cD galaxy.}
\label{fig:f5}
\end{center}
\end{figure}

\subsubsection{Radio spectra}
\label{sec.rad-spec}

The total flux densities were determined from integration in polygons with measured in areas encompassing the remnant radio galaxy, {\it i.e.}, we determined the surface-brightness and divided it by the number of pixels in the polygon at each frequency.  These results are presented in Table~\ref{tab.rad-spectra} along with the total intensity spectra of the remnant radio galaxy, and the southern cD galaxy is shown in Figure~\ref{fig:f5}.  A 2.8$\sigma$ upper limit at the 74\,MHz \citep[VLSS at 80$^{\prime\prime}$ angular resolution;][]{2007AJ....134.1245C} and the peak flux density from our 150\,MHz band data for the southern cD galaxy are also reported.  We see that all these radio measurements, within the error bars, fall nearly along with a `straight' power law, $\alpha$ = $-$1.2 $\pm$0.2 with some hints of energy losses by synchrotron cooling for the dominant southern cD galaxy.  This provides confidence in our data calibration and its reduction methodologies.

The remnant radio galaxy is not detected in our 150\,MHz band short observation, and is barely detected using LOFAR at 153~MHz (see also Sec.~\ref{sec.radio-relic}).
However in the 1050--1450\,MHz band short observation, we see hints of radio emission from the remnant radio galaxy.  We made a low-resolution, 15$^{\prime\prime}$ image at the 1050--1450\,MHz band and report peak flux density for the radio emission from the remnant radio galaxy in Figure~\ref{fig:f5}. 
Since this 1050--1450\,MHz band measurement comes from a short $\lesssim$0.9~hr duration observation, has relatively high \textsc{rms} noise (see also Sec.~\ref{sec.results}), and the observation does not sample similar ($u,v$)-spacings as the rest of the low-frequency observations, we henceforth no longer include it in our spectral index analysis.
It appears that the integrated spectrum for the remnant radio galaxy cannot be represented by a single power law.  The segment of the spectrum with the best data is $<$ 1250~MHz, and the low-frequency spectral index $\alpha$(240--610\,MHz) =  $-$1.4 $\pm$0.2.  Our data cannot rule out the possibility of spectral shape to be due to two populations of the relativistic electrons: a low-frequency component characterized by $\alpha$(240--610 MHz) $\approx$ $-$1.4 and a second component having $\alpha$($>$ 610\,MHz) $<$ $-$1.4 with a cutoff or a spectral break between 400 and 1250\,MHz.

\begin{figure}[b]
\begin{center}
\begin{tabular}{c}
\includegraphics[width=8.45cm]{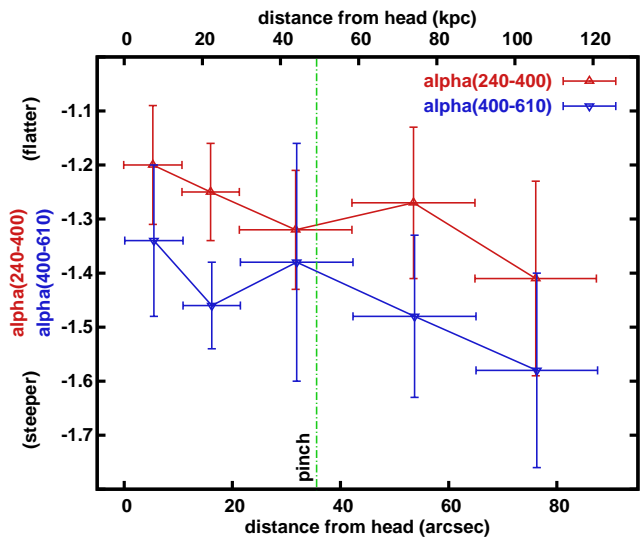}
\end{tabular}
\caption{The plot shows the radio spectra between the 240 MHz and 250--500 MHz band data sets (red upper triangles) and between the 250--500 MHz band and 610 MHz band data sets (blue lower triangles) as a function of distance from the head.  The location of the pinch is shown as a vertical line.}
\label{fig:f6}
\end{center}
\end{figure}

\begin{figure}[t]
\begin{center}
\begin{tabular}{c}
\includegraphics[width=8.45cm]{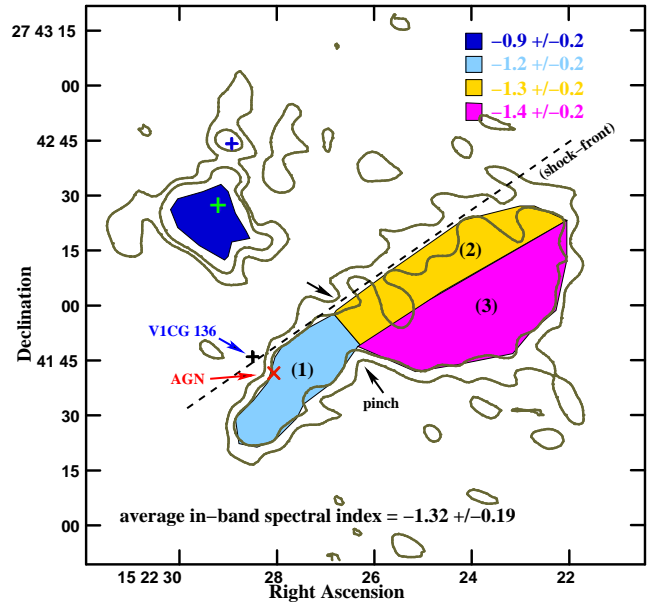}
\end{tabular}
\caption{The averaged in-band spectral index image showing spectra between 300 MHz and 500 MHz using 250--500 MHz band data.  The total intensity surface-brightness contours are at 0.23 and 0.46 mJy~beam$^{-1}$ from the 300--500 MHz radio image (Figure~\ref{fig:f4}, top-right panel).  We used three irregular polygon regions, marked in the image, and determined the mean (and rms) spectral indices that are denoted at the upper-right corner of the image.  The averaged integrated in-band spectral index for the remnant radio galaxy is denoted at the lower-left corner of the image.  The cross marks correspond to the positions of two cD galaxies and the position of the optical host galaxy, discussed in Sec.~\ref{sec.Xray-two-cDs} and \ref{sec.radio-relic}.  The dashed line indicates the best-fitting position of the X-ray surface-brightness edge discussed in Sec.~\ref{sec.xray-prop}.}
\label{fig:f7}
\end{center}
\end{figure}

\begin{figure*}[ht]
\begin{center}
\begin{tabular}{c}
\includegraphics[width=14.4cm]{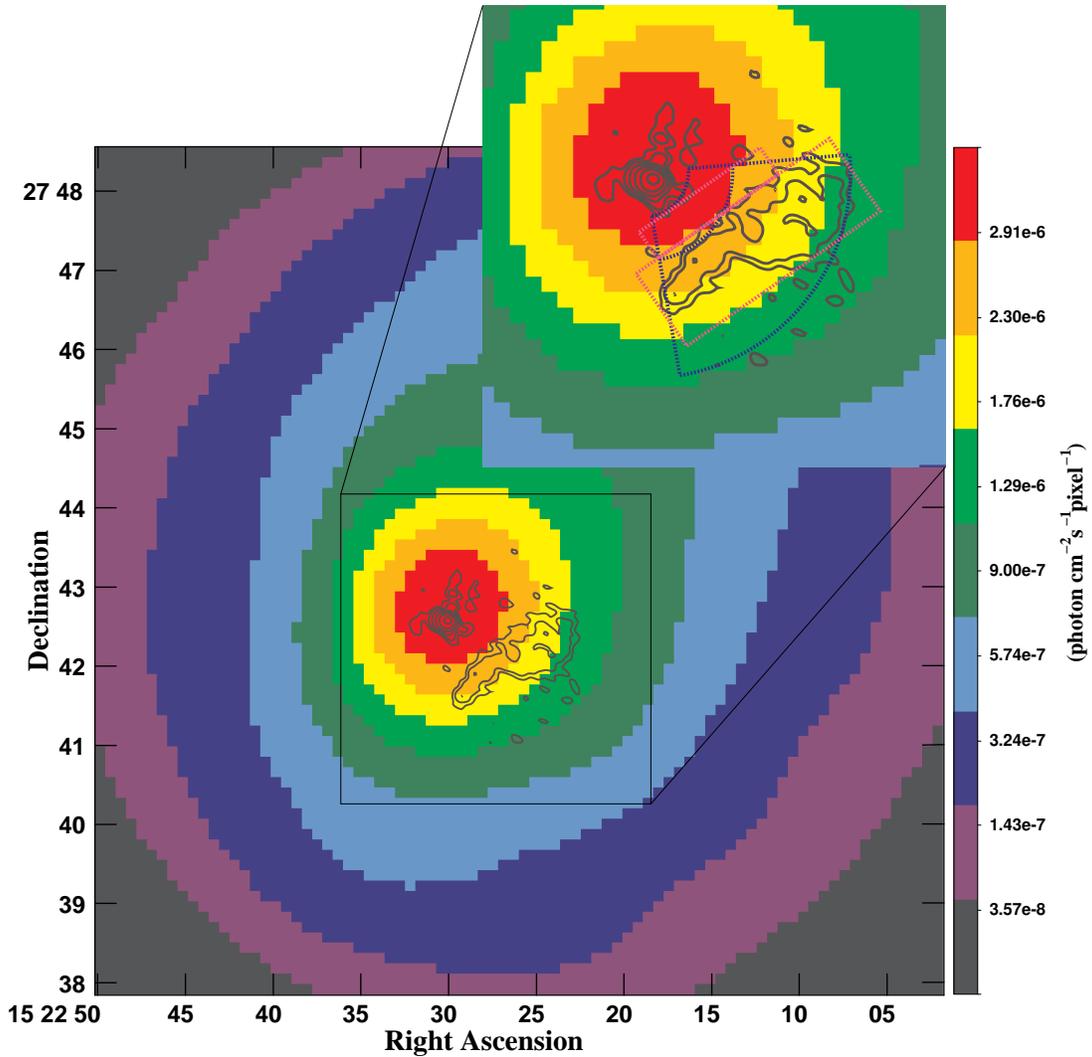}
\end{tabular}
\caption{Background-subtracted and exposure-corrected \textit{Chandra} image at $\sim$4$^{\prime\prime}$ angular resolution in the broad (0.5--7.0 keV) energy band.  The color bar shows the surface-brightness, in photon~cm$^{-2}$~s$^{-1}$~pixels$^{-1}$.  The black surface-brightness contours are displayed using the 250--500 MHz band image.  The inset shows the enlargement of the inner region.  Also shown are two sets of sectors (in cyan and blue colors) that were used for the surface-brightness and temperature measurements.}
\label{fig:f8}
\end{center}
\end{figure*}

The spectral index measurements between the 240 MHz band and 250--500 MHz band data sets, and between the 250--500 MHz band and 610 band MHz band data sets show that the radio spectrum steepens with distance from the head, from $-$1.2 $\pm$0.1 to $-$1.4 $\pm$0.2, and from $-$1.3 $\pm$0.1 to $-$1.6 $\pm$0.2, respectively, due to synchrotron and inverse Compton losses (Figure~\ref{fig:f6}).  The location after the pinch, where the radio plasma from the AGN connects to the remnant radio galaxy, the spectrum, between the 240 MHz band and 250--500 MHz band data, remains flattened, $\alpha$ = $-$1.3 $\pm$0.1.

The wide-bandwith data sets were imaged to make multiscale multifrequency synthesis images, which provides images at center frequencies and the corresponding in-band spectral index maps (see also Sec.~\ref{uGMRT-data}).
The in-band spectral index map averaged over a region of interest, to increase signal-to-noise ratio, for the 250--500~MHz band data is shown in Figure~\ref{fig:f7}.  We divided the remnant radio galaxy into three distinct regions: (1) the high-surface-brightness narrower southeastern part 
ahead of the pinch (in cyan).  The latter low-surface-brightness broader northwestern part is divided into two; (2) the northeastern region (in yellow), and (3) the southwestern far end region (in magenta).
The size of the regions is chosen such that the signal in the total intensity image is (approximately) more than five times the \textsc{rms} noise.  The southern cD galaxy has an in-band spectral index of $-$0.9 $\pm$0.2.  Similarly, the spectra of three regions of the remnant radio galaxy are $-$1.2 $\pm$0.2, $-$1.3 $\pm$0.2, and $-$1.4 $\pm$0.2 for regions (1), (2), and (3) respectively.  This is consistent with the averaged in-band spectral index of $-$1.3 $\pm$0.2 within the error bars for the remnant radio galaxy.  It is also consistent with the crude estimates of two 50 MHz subband images, 300--350 MHz and 450--500~MHz, at two extreme ends of the 250--500 MHz band data.
The region ahead of the pinch (in cyan) and the northeastern region close to two cD galaxies (in yellow) have comparable spectral indices.
It appears that the high-surface-brightness narrower region (in cyan) ahead of the pinch, is flatter than the two regions, together forming low-surface-brightness broader northwestern region (in yellow and magenta) after the pinch, consistent with our result discussed above.  Furthermore, the area between the two regions of the broader northwestern part of the source, {\it i.e.}, the southwestern far-end region (in magenta) have a significantly steeper spectrum than the northeastern region close to two cD galaxies (in yellow).

\subsection{ICM gas properties}
\label{sec.xray-prop}

\begin{figure}[t]
\begin{center}
\begin{tabular}{c}
\includegraphics[width=8.4cm]{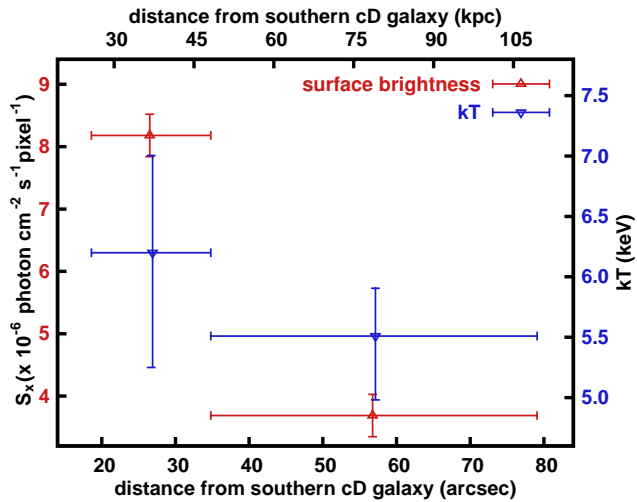}
\end{tabular}
\caption{Plot of the radial surface-brightness profile and the temperature profile for the two red sectors (shown in Figure~\ref{fig:f8}) as a function of distance from the southern cD galaxy.  The error bars represent 90\% confidence intervals.}
\label{fig:f9}
\end{center}
\end{figure}

\citet{Chatzikosetal2006} reported that the cluster contains gas with an extremely wide range of temperatures and a wealth of structure.  They also reported a discontinuity at $\sim$30$^{\prime\prime}$ ($\simeq$ 41.5~kpc) in the southeast region from the southern cD galaxy.  It is at nearly this same distance where we report an enhancement of radio surface-brightness, toward the two cD galaxies, showing a sharp feature.  We aim to search an X-ray surface-brightness edge, {\it i.e.}, the shock front that is coincident with the northeastern sharp edge toward two cD galaxies of the remnant radio galaxy.
In order to understand the gas physics at this region of the cluster, we have extracted the surface-brightness profile from two box-shaped regions, one encompassing the remnant radio galaxy (called ``outside" here) and the other adjacent to it toward cD galaxies (called ``inside" here) as shown in Figure~\ref{fig:f8} (in magenta).  We extracted the surface-brightnesses and the spectra from these regions, which are displayed in Figure~\ref{fig:f9}.  A single temperature model with Galactic absorption and the abundance fixed provided a good fit to the spectrum.

We noticed the X-ray surface-brightness edge, where the surface-brightness drops (see Figures~\ref{fig:f8} and \ref{fig:f9}) and therefore implies a discontinuity, or jump, in the gas density.  The surface-brightnesses are $S_{\rm x}$(inside) and $S_{\rm x}$(outside) $\approx$ 8.2 $\pm$0.3 $\times$ 10$^{-6}$ counts~s$^{-1}$~arcsec$^{-2}$ and 3.7 $\pm$0.3 $\times$ 10$^{-6}$ counts~s$^{-1}$~arcsec$^{-2}$, respectively.  Consequently, our modeled densities in the two regions, $\rho$(inside) and $\rho$(outside) are approximately 2.9$^{+0.2}_{-0.1}$ $\times$ 10$^{-3}$ cm$^{-3}$ and 1.9$^{+0.2}_{-0.2}$ $\times$ 10$^{-3}$ cm$^{-3}$, respectively.
Our corresponding derived temperatures for the two regions, $kT_{\rm (inside)}$ and $kT_{\rm (outside)}$ are approximately 6.1$^{+0.9}_{-0.8}$~keV ($\chi^2$ = 0.93 for 221 degrees of freedom) and 5.5$^{+0.5}_{-0.4}$~keV ($\chi^2$ = 1.08 for 462 degrees of freedom), respectively.
We also repeated this exercise for a pie-circular aperture (shown in blue in Figure~\ref{fig:f8}) centered on the southern cD galaxy and lying toward the radio emission from the remnant radio galaxy with two regions containing the radio emission from the remnant radio galaxy and the smaller adjacent region toward the cD galaxy.  The quantitative estimates from this are nearly identical and the fit statistics are at 90\% confidence intervals.  Although with large error bars, these density and temperature measurements are hinting that the surface-brightness edge toward the two cD galaxies is possibly a shock front.
Consequently, the observed density jump and the temperature jump across the edge, $\rho_{\rm (inside)}$/$\rho_{\rm (outside)}$ = 1.5$^{+0.2}_{-0.2}$ and $kT_{\rm (inside)}$/$kT_{\rm (outside)}$ = 1.1$^{+0.7}_{-0.5}$ suggest the Mach numbers, $M$ = 1.3$^{+0.2}_{-0.2}$ and 1.1$^{+0.4}_{-0.3}$, respectively under the Rankine-Hugoniot shock conditions for the intracluster gas, $\gamma$ = 5/3, which are all obviously in agreement, are within the error bars of \citet{Chatzikosetal2006}.
We also believe that a smaller measure for the temperature jump is due to the contribution of cool thermal emission from the core to the adjacent region toward two cD galaxies, $kT_{\rm (inside)}$.
Note, it is equally possible that the surface-brightness edge is a cold front.  Here, we assume it to be a shock front because of consistencies between our measurements and those of \citet{Chatzikosetal2006} and \citet{Markevitchetal1999}.

\section{Discussion}
\label{sec.discuss}

In passing, we note that we do not detect the radio halo of A2065 in our current $\approx 8^{\prime\prime}$ resolution 250--500 MHz band image. This is likely to be due to the fairly short duration of our observing run (just $\approx 1.3$~hr), which results in poor ($u,v$)-coverage at the short spacings that are critical to detect $\approx 10^\prime$-sized structures like the radio halo \citep[e.g.][]{Farnsworthetal}.  However, the large fractional bandwidth at 250--500 MHz band implies that it has excellent ($u,v$)-coverage at the intermediate and long baselines that are needed to accurately map the arcminute-sized structures like our target, the remnant radio galaxy (angular size~$\approx 1^\prime \times 2^\prime$).
In addition, we have performed deep \textsc{clean}, where we have used the multiscale multifrequency synthesis process of combining data from multiple spectral channels onto the same spatial-frequency grid during imaging to take advantage of the increased ($u,v$)-coverage and imaging sensitivity, and hence we do not foresee any loss of flux density that could be linked to the remnant radio galaxy in the data presented above.
Below we discuss the nature of this newly discovered remnant radio galaxy in A2065.

As the remnant radio galaxy ages, its electrons will lose so much energy that their spectra will eventually steepen.  This steepening would reach to a point where they do not emit significant radiation at high frequencies, instead, the radio emission may only be observable at low frequencies.
However, \citet{Brienzaetal} argues that in addition to the aged population of remnant radio galaxies, there should be younger population as well in which lobes have not had time to steepen over the observed frequency range \citep[e.g., 3C\,28 and MIDAS J230304$-$323228 in][respectively]{Ferettietal1984,Quicietal}.
It is possible that the remnant phase of a radio galaxy is governed by a Sedov-like expansion, which will cause the dimming of radio emission due to a decrease in the magnetic field strength and a decrease in particle energies due to adiabatic expansion losses.  Unfortunately, the models of the spectra of remnant radio galaxy are highly degenerate, and the modeling of an individual remnant radio galaxy cannot constrain the remnant phase of radio galaxy \citep{KaiserandCotter}.
Therefore, a possible way to constrain the remnant phase is via a statistical approach, and sadly, due to their small number, the physics of remnant radio galaxies remains poorly understood.
Of course, alongside their must exist a significant population of remnant radio galaxies, which show the absence of detectable radio emission in many clusters associated with merger activity \citep[see also][]{Godfrey2017}.
Thus, searching for low-surface-brightness profiles, say $\lesssim$ 50~mJy~arcmin$^{-2}$, with an absence of radio cores and hotspots, and with no well-defined radio morphologies, is a possible way to identify remnant radio sources in large survey images.

Radio morphology of the remnant radio galaxy in A2065 shows that it is an elongated, bar-shaped structure (52$^{\prime\prime}$ $\times$ 110$^{\prime\prime}$ $\simeq$ 72 $\times$ 152 kpc$^2$), which is buried within a radio halo emission of size $\lesssim$ 1~Mpc in size \citep{Drabentetal} and \citep{Farnsworthetal}.
The remnant radio galaxy is located in the near vicinity of the southern cD galaxy; the northeastern boundary of the remnant radio galaxy is merely at a distance of $\sim$33$^{\prime\prime}$ ($\simeq$ 45.7~kpc) from the southern cD galaxy.
The possible radio core of the remnant radio galaxy is coincident with the WISEA J152228.01$+$274141.3 source that is 7\farcs8 away from the optical galaxy V1CG\,136.
Our spectral index measurements show that the radio spectrum steepens with distance from the possible radio core from $\alpha$ = $-$1.2 $\pm$0.1 to $\alpha$ = $-$1.4 $\pm$0.2 in our low-frequency (between the 240 MHz band and the 250--500 MHz band) spectra, and from $\alpha$ = $-$1.3 $\pm$0.1 to $\alpha$ = $-$1.6 $\pm$0.2 in our high-frequency (between the 250--500 MHz band and the 610 MHz band) spectra (Figure~\ref{fig:f6}), as is expected for synchrotron and inverse Compton losses.  We also find evidence for spectral steepening across the remnant radio galaxy in our in-band spectral index measurements, from $\alpha$ = $-$1.3 $\pm$0.2 to $\alpha$ = $-$1.4 $\pm$0.2 (Figure~\ref{fig:f7}), in the direction away from the two cD galaxies.
Our X-ray measurements show a surface-brightness edge that is coincident with the northeastern boundary of the remnant radio galaxy and its nature (shock or cold front) is not clear from the data.  If it is a shock front, it is rather weak, with $M$ being between 1.1 and 1.3; however it is unclear if an (inward?) traveling shock is responsible for spectral steepening across the remnant radio galaxy.
Below we discuss the possible progenitor of the remnant radio galaxy in A2065.

\subsection{Remnant source diagnostic}

Remnant radio sources are old, and are bound by model-dependent physical parameters because of the required assumptions of volume-filling factors, etc.  We use the parameter $N$, which is the number of relativistic electrons found by integrating the energy spectrum of radiating electrons between Lorentz factors of 1000 and 10,000 for a typical equipartition magnetic field of 5~$\mu$G in the 2--200\,MHz band for the remnant radio galaxy.
The number distribution of relativistic electrons drops rapidly with energy for a steep spectrum remnant radio galaxy.
Hence the number of low-energy electrons is a good measure of the total acceleration of relativistic electrons during the lifetime of $\sim$10$^8$~yr for the small redshift of the source, and indicates the entire history of the source.

We thus determine $N$ $\approx$ 2.1 $\times$ 10$^{60}$ and the total energy stored in the magnetic field $\approx$ 1.1 $\times$ 10$^{57}$~erg assuming cylindrical geometry for the size of the remnant radio galaxy \citep{Miley1980,Pacholczyk1970}.  Here, we have assumed the volume filling factor is unity, and hence the volume and the number of relativistic electrons $N$ are thus the upper limits.  The emitting plasma will most likely occupy a smaller volume, leading to larger energy densities and therefore larger values of the magnetic field strength and will require a smaller number of electrons to produce the observed luminosity.  These basic energetics are similar to the Coma radio relic source 1253$+$275 \citep{Giovanninietal}.  Moreover, the number of relativistic electrons, $N$, is similar to that of the typical FR\,I radio galaxy 3C\,31 but two orders of magnitude smaller than the FR\,II radio galaxy Cygnus\,A \citep{Harrisetal}.  This suggests that the FR\,I radio galaxy is possibly the progenitor of the remnant radio galaxy in A2065.

Thus, our high-resolution, high-sensitivity radio image at the 250--500 MHz band along with the presence of the WISEA J152228.01$+$274141.3 source
suggest that the source is a wide-angle-tailed radio source (see also Sec.~\ref{sec.radio-relic}).
Furthermore, the position of the WISEA J152228.01$+$274141.3 source exactly coincides with the likely radio core position of the wide-angle-tailed radio source.  The source diagnostic based on the number of low-energy relativistic electrons also suggests that the remnant radio galaxy is fed by the small radio tail of the wide-angle-tailed FR\,I radio galaxy.
If this proposition is correct, the two oppositely directed radio jets emanating from the apex of the WISEA J152228.01$+$274141.3 source initially traverse toward the northwest and southeast.  As the galaxy plows through the dense ICM, these jets traversing in opposite directions form a trail after a sharp bend in the southeastern jet due to interaction with the ICM.  The jets probably overlap and twist forming a pinch and then expand, and the jet becomes weaker \citep[e.g., NGC\,4869;][]{Lal2020a}.  The sharpness in surface-brightness toward the northeast, {\it i.e.}, the side facing two cD galaxies and diffuse extensions toward the far southwest side, along with corresponding spectral gradients, suggests the passage of an (inward?) traveling shock front.

\section{Summary and conclusions}
\label{sec.sum-conc}

The remnant radio source seems to trace AGN radio plasma that has somehow been reenergised through processes in the ICM.
It is possible that, similar to radio relic sources, remnant radio sources are also believed to be tracers of merger shocks, unfortunately, only a few clusters have shown to possess corresponding features in hot X-ray gas.  The remnant radio galaxy in A2065 may belong to this (rare) class possessing an X-ray surface-brightness edge across the remnant radio galaxy, indicating a possible connection with the shock front.
Using our deep high-resolution low-frequency observations, we discovered a bar-shaped remnant radio galaxy, whose size is $\approx$ 52$^{\prime\prime}$ $\times$ 110$^{\prime\prime}$ (= 72 $\times$ 152 kpc$^2$), remnant radio galaxy in the massive unequal-mass-merger galaxy cluster A2065 at $z$ = 0.072.
It has a steep spectral index $\alpha$(240--610\,MHz) = $-$1.4 $\pm$0.2 and $\alpha$($>$ 610\,MHz) $<$ $-$1.4 $\pm$0.2, corresponding to the line representing the best-fitting regression to the radio data.
The \textit{Chandra} data, and the upgraded and ``classic" GMRT data, show
(i) the presence of an X-ray surface-brightness edge attributed to a shock front, though a possibility of cold front cannot be ruled out, at the remnant radio galaxy's edge toward the two cD galaxies, (ii) the tentative flattening of the radio spectral index, in our 250--500 MHz in-band data, of the remnant radio galaxy at the near end of the X-ray surface-brightness edge, and (iii) possibly a wide-angle-tailed, FR\,I radio galaxy is the progenitor of the remnant radio galaxy.
Thus, the remnant radio galaxy has possibly been reinvigorated by the passage of the shock front and adiabatic compression, and shows the expected change in radio emission.  We also suggest that the remnant radio galaxy was seeded by the AGN during its past active phase, and has been hosted by the WISEA J152228.01$+$274141.3 source, demonstrating the connection between AGNs and remnant radio galaxy, similar to the connection between AGNs and radio relic sources \citep[e.g., A3411-3412,][]{vanWeerenetal}.

Although low-frequency observations are starting to reveal more and more of these type of sources, presently there are a very few known remnant radio systems; clearly an outstanding difficulty is the small sample size.
Although the number of known remnant radio sources is increasing, better data are necessary for several of them to allow a detailed study. Fortunately, forthcoming radio surveys using the LOFAR, MeerKAT, Square Kilometre Array, etc., will greatly increase the sample size, along with deep multifrequency radio data, e.g., using the upgraded GMRT, which will constrain the physics and nature of poorly understood remnant radio sources.
 We are undertaking an upgraded GMRT study at the 125--250 MHz, 250--500 MHz, and 550--850 MHz bands of several clusters in order to detect such ultra-steep-spectrum sources, map radio halo emission, and provide exact flux density estimates, models, and statistics.

\section*{Acknowledgments}

I would like to thank the anonymous referee for helpful remarks.
D.V.L. acknowledges the support of the Department of Atomic Energy, Government of India, under project No. 12-R\&D-TFR-5.02-0700, and thanks the staff of the GMRT who made these observations possible. The GMRT is run by the National Centre for Radio Astrophysics of the Tata Institute of Fundamental Research.
This research made use of the radio astronomical database GalaxyClusters.com, maintained by the Observatory of Hamburg.
This research has made use of the NED, which is operated by the Jet Propulsion Laboratory, Caltech, under contract with the NASA, and NASA's Astrophysics Data System.

Facilities: \facility{Chandra, GMRT}

Data Availability: The GMRT and \textit{Chandra} data underlying this article are available via the GMRT Online Archive, {naps.ncra.tifr.res.in/goa} and the \textit{Chandra} Data Archive, {cxc.harvard.edu/cda}.
All data analyses packages used in this work are publicly available.


\begin{thebibliography}{}

\bibitem[Arnaud(1996)]{Arnaud1996} Arnaud K. A., 1996, in Jacoby G. H., Barnes J., eds, ASP Conf. Ser. Vol.101, Astronomical Data Analysis Software and Systems V. Astron. Soc. Pac., San Francisco, p. 17
%

\bibitem[Brienza et~al.(2017)]{Brienzaetal} Brienza, M., Godfrey, L. E. H., Morganti, R., et~al. 2017, A\&A 606, A98

\bibitem[Chatzikos, Sarazin \& Kempner(2006)]{Chatzikosetal2006} Chatzikos, M., Sarazin, C. L., \& Kempner, J. C., 2006, ApJ, 643, 751

\bibitem[Cohen et~al.(2007)]{2007AJ....134.1245C} Cohen A.S., Lane W.M., Cotton W.D., et~al. 2007, AJ, 134, 1245

\bibitem[Dickey \& Lockman(1990)]{DickeyLockman} Dickey, J. M., \& Lockman, F. J., 1990, ARA\&A, 28, 215

\bibitem[Drabent et~al.(2017)]{Drabentetal} Drabent, A., Hoeft, M., Br\"uggen, M., et~al., 2017, in ``Diffuse synchrotron emission in clusters of galaxies - what's next?", {https://indico.cern.ch/event/617524/contributions/2743186/}

\bibitem[Farnsworth et~al.(2013)]{Farnsworthetal} Farnsworth, D.,  Rudnick, L.,  Brown, S.,  Brunetti, G. 2013, ApJ, 779, 189

\bibitem[Feretti et~al.(1984)]{Ferettietal1984} Feretti L., Gioia I. M., Giovannini G., Gregorini L., Padrielli L. 1984, A\&A, 139, 50

\bibitem[Giovannini \& Feretti(2002)]{GiovanniniandFeretti2002} Giovannini, G. \& Feretti, L. 2002 in Merging Processes in Galaxy Clusters, eds. L. Feretti, I. M. Gioia, \& G. Giovannini, ASSL, Kluwer Ac. Publish., p. 197

\bibitem[Giovannini, Feretti \& Stanghellini(1991)]{Giovanninietal} Giovannini, G., Feretti, L., Stanghellini, C., 1991, A\&A, 252, 528

%
\bibitem[Godfrey, Morganti \& Brienza(2017)]{Godfrey2017} Godfrey, L. E. H., Morganti, R. \& Brienza, M. 2017, MNRAS, 471, 891

\bibitem[Griffin, Dai \& Kochanek(2014)]{2014ApJ...795L..21G} Griffin, R. D., Dai, X., \& Kochanek, C. S., 2014, ApJL, 795, L21

\bibitem[Harris et~al.(1993)]{Harrisetal} Harris, D. E., Stern, C. P., Willis, A. G., \& Dewdney, P. E., 1993, AJ 105, 769

\bibitem[Hoeft et~al.(2004)]{Hoeftetal2004} Hoeft, M., Br\"uggen, M., Yepes, G., et~al. 2004, MNRAS, 347, 389

%
\bibitem[Kaiser \& Cotter(i2002)]{KaiserandCotter} Kaiser, C. R. \& Cotter, G. 2002, MNRAS, 336, 649

\bibitem[Kempner et~al.(2004)]{Kempneretal} Kempner, J. C., Blanton, E. L., Clarke, T. E. et~al. in proc. of The Riddle of Cooling Flows in Galaxies and Clusters of Galaxies, p. 335

\bibitem[Lal(2020)]{Lal2020} Lal, D. V., 2020, ApJS, 250, 22

\bibitem[Lal(2020a)]{Lal2020a} Lal, D. V., 2020, AJ, 160, 161

\bibitem[Lal et~al.(2019)]{Laletal} Lal, D. V., Sebastian, B., Cheung, C. C., Rao, A. P. 2019, AJ, 157, 195

\bibitem[Markevitch, Sarazin \& Vikhlinin(1999)]{Markevitchetal1999} Markevitch, M., Sarazin, C. L., \& Vikhlinin, A., 1999, ApJ, 521, 526

\bibitem[Miley(1980)]{Miley1980} Miley, G. K., 1980, ARA\&A, 18, 165

\bibitem[Pacholczyk(1970)]{Pacholczyk1970}Pacholczyk, A. G., 1970, Radio Astrophysics (San Francisco: W. E. Freeman \& Co.)

%

\bibitem[Parma et~al.(2007)]{Parmaetal2007} Parma, P., Murgia, M., de Ruiter, H. R., et~al. 2007, A\&A, 470, 875
%
\bibitem[Postman, Geller \& Huchra(1988)]{Postmanetal}Postman M., Geller, M. J., \& Huchra, J. P. 1988, AJ, 95 267

\bibitem[Quici et~al.(2021)]{Quicietal} Quici, B., Hurley-Walker, N., Seymour, N., et~al. 2021, PASA, 38, 8


\bibitem[Struble \& Rood(1999)]{1999ApJS..125...35S} Struble, M. F., \& Rood, H. J., 1999, ApJS, 125, 35

%
\bibitem[van~Weeren et~al.(2019)]{vanWeerenetal2019} van~Weeren, R. J., de~Gasperin, F., Akamatsu, H., et~al. 2019, Space Sci. Rev., 215, 16

\bibitem[van~Weeren et~al.(2017)]{vanWeerenetal} van~Weeren, R. J., Andrade-Santos, F., Dawson, W. A., et~al. 2017, Nat. Astron. 1, 5


\end{thebibliography}
\end{document}